\documentclass[aps, prb, preprint, preprintnumbers, superscriptaddress, showpacs, amssymb]{revtex4-1}

\usepackage{graphicx}% Include figure files
\usepackage{dcolumn}% Align table columns on decimal point
\begin{document}

% Use the \preprint command to place your local institutional report
% number in the upper righthand corner of the title page in preprint mode.
% Multiple \preprint commands are allowed.
% Use the 'preprintnumbers' class option to override journal defaults
% to display numbers if necessary
\preprint{Ver. 4p2JPSJ}¡¡

%Title of paper
\title{Pressure-Induced Antiferromagnetic Transition and Phase Diagram in FeSe}

% repeat the \author .. \affiliation  etc. as needed
% \email, \thanks, \homepage, \altaffiliation all apply to the current
% author. Explanatory text should go in the []'s, actual e-mail
% address or url should go in the {}'s for \email and \homepage.
% Please use the appropriate macro foreach each type of information

% \affiliation command applies to all authors since the last
% \affiliation command. The \affiliation command should follow the
% other information
% \affiliation can be followed by \email, \homepage, \thanks as well.
\author{Taichi Terashima}
\author{Naoki Kikugawa}
\affiliation{National Institute for Materials Science, Tsukuba, Ibaraki 305-0003, Japan}
\author{Shigeru Kasahara}
\author{Tatsuya Watashige}
\affiliation{Department of Physics, Kyoto University, Kyoto 606-8502, Japan}
\author{Takasada Shibauchi}
\affiliation{Department of Physics, Kyoto University, Kyoto 606-8502, Japan}
\affiliation{Department of Advanced Materials Science, University of Tokyo, Chiba 277-8561, Japan}
\author{Yuji Matsuda}
\affiliation{Department of Physics, Kyoto University, Kyoto 606-8502, Japan}
\author{Thomas Wolf}
\author{Anna E. B\"ohmer}
\author{Fr\'ed\'eric Hardy}
\author{Christoph Meingast}
\author{Hilbert v. L\"ohneysen}
\affiliation{Institute of Solid State Physics (IFP), Karlsruhe Institute of Technology, D-76021 Karlsruhe, Germany}
\author{Shinya Uji}
\affiliation{National Institute for Materials Science, Tsukuba, Ibaraki 305-0003, Japan}

%Collaboration name if desired (requires use of superscriptaddress
%option in \documentclass). \noaffiliation is required (may also be
%used with the \author command).
%\collaboration can be followed by \email, \homepage, \thanks as well.
%\collaboration{}
%\noaffiliation

\date{\today}
\begin{abstract}
We report measurements of resistance and ac magnetic susceptibility on FeSe single crystals under high pressure up to 27.2 kbar.
The structural phase transition is quickly suppressed with pressure, and the associated anomaly is not seen above $\sim$18 kbar.
The superconducting transition temperature evolves nonmonotonically with pressure, showing a minimum at $\sim12$ kbar.
We find another anomaly at 21.2 K at 11.6 kbar.
This anomaly most likely corresponds to the antiferromagnetic phase transition found in $\mu$SR measurements [M. Bendele \textit{et al.}, Phys. Rev. Lett. \textbf{104}, 087003 (2010)].
The antiferromagnetic and superconducting transition temperatures both increase with pressure up to $\sim25$ kbar and then level off.
The width of the superconducting transition anomalously broadens in the pressure range where the antiferromagnetism coexists.
\end{abstract}

% insert suggested PACS numbers in braces on next line
%\pacs{74.70.Xa, 74.62.Fj, 74.25.Dw}
% insert suggested keywords - APS authors don't need to do this
%\keywords{}

%\maketitle must follow title, authors, abstract, \pacs, and \keywords
\maketitle

% body of paper here - Use proper section commands
% References should be done using the \cite, \ref, and \label commands
%\section{}
% Put \label in argument of \section for cross-referencing
%\section{\label{}}
%\subsection{}
%\subsubsection{}

% If in two-column mode, this environment will change to single-column
% format so that long equations can be displayed. Use
% sparingly.
%\begin{widetext}
% put long equation here
%\end{widetext}

\newcommand{\ud}{\mathrm{d}}
\def\degree{\kern-.2em\r{}\kern-.3em}

The superconductivity in FeSe has attracted growing attention since its discovery.\cite{Hsu08PNAS}
FeSe is unique among iron-based superconductors.
In contrast to the iron-pnictide parent compounds such as LaFeAsO (Ref.~\onlinecite{Kamihara08JACS}) and BaFe$_2$As$_2$,\cite{Rotter08PRB} FeSe undergoes a structural transition at $T_s \sim90$ K but no magnetic transition at ambient pressure.
It becomes superconducting below $T_c \sim 8$ K.\cite{Hsu08PNAS}
The transition temperature can be enhanced substantially by the application of pressure,\cite{Mizuguchi08APL} the superconducting onset temperature reaching $\sim37$ K at $P \sim89$ kbar.\cite{Medvedev09Nmat}
Moreover, single-layer FeSe films may have still higher transition temperatures.\cite{Wang12CPL}
Recent breakthroughs in the crystal growth (Refs.~\onlinecite{Chareev13CrystEngComm, Bohmer13PRB}) have led to a flurry of research activities, revealing more and more peculiarities of FeSe.\cite{Maletz14PRB, Shimojima14PRB, Huynh14PRB, Terashima14PRB, Kasahara14PNAS, Baek14nmat, Nakayama14PRL, Bohmer15PRL, Audouard15EPL}
Quantum oscillation and angle-resolved photoemission spectroscopy (ARPES) measurements have found an unexpectedly shrunk Fermi surface.\cite{Maletz14PRB,Terashima14PRB}
Some other ARPES measurements have found a splitting of the $d_{xz}$ and $d_{yz}$ bands below $\sim T_s$,\cite{Shimojima14PRB, Nakayama14PRL} suggesting that the structural transition is an electronic nematic order driven by orbital degrees of freedom, which is in accord with a conclusion from thermodynamic and NMR measurements.\cite{Bohmer13PRB, Baek14nmat, Bohmer15PRL}
Thermal conductivity measurements have found a phase transition within the superconducting phase.\cite{Kasahara14PNAS}
It has also been argued that because of small Fermi energies the superconductivity in FeSe may be close to a Bardeen-Cooper-Schrieffer (BCS)--Bose-Einstein-condensation (BEC) crossover.\cite{Kasahara14PNAS}

The absence of the antiferromagnetic order at ambient pressure may be at the heart of the FeSe enigma.
Recent theoretical studies have suggested that the absence is due to the competition between spin fluctuations with different wave vectors.\cite{Lischner15PRB, Glasbrenner15condmat}
Experimentally, a pressure-induced antiferromagnetic order was already suggested by early NMR measurements (Ref.~\onlinecite{Imai09PRL}) and has been confirmed by recent $\mu$SR measurements.\cite{Bendele10PRL, Bendele12PRB}
However, there is so far hardly any evidence from macroscopic measurements such as resistivity or magnetization under high pressure \cite{Mizuguchi08APL, Braithwaite09JPCM, Masaki09JPSJ, Medvedev09Nmat, Margadonna09PRB, Miyoshi09JPSJ, Okabe10PRB, Miyoshi14JPSJ}: for the pressure range of the present study, only Ref.~\onlinecite{Masaki09JPSJ} observed kinks in the temperature dependence of resistivity of polycrystalline samples at pressures above 28.5 kbar and speculated that they might be of a magnetic origin.
In addition, to our knowledge, no previous studies observed both of the structural and antiferromagnetic transitions in the same sample under high pressure.

In this paper, we report resistance and ac magnetic susceptibility measurements on FeSe under high pressure up to $P$ = 27.2 kbar.
Our resistance measurements clearly show anomalies that are most likely associated with the antiferromagnetic transition.
This corroborates the appearance of a static long-range order under high pressure.
We construct a phase diagram composed of the three phase transitions, i.e., the superconducting, antiferromagnetic, and structural ones, which displays an intriguing  interplay between those orders.  

We performed four-contact electrical resistance measurements on high-quality single crystals of FeSe prepared by a chemical vapor transport method (Ref.~\onlinecite{Bohmer13PRB}) down to helium temperatures at pressures up to 27.2 kbar.
The electrical contacts were spot-welded.
The low-frequency ac current ($f$ = 13 Hz) was applied in the $ab$ plane.
The current density was roughly in a range 6 -- 8 A/cm$^2$.
Piston-cylinder type pressure cells made of NiCrAl alloy (C\&T Factory, Tokyo) were used.\cite{Uwatoko02JPCM}
The pressure transmitting medium was Daphne 7474 (Idemitsu Kosan, Tokyo), which remains liquid up to 37 kbar at room temperature and assures highly hydrostatic pressure generation in the investigated pressure range.\cite{Murata08RSI} 
The pressure was determined from the resistance variation of calibrated manganin wires.
We also performed ac magnetic susceptibility measurements under high pressure, where the ac excitation field ($B_{ac}$ $\sim 2 \times 10^{-5}$ T, $f$ = 67 Hz) was applied approximately parallel to the $ab$ plane.

\begin{figure}
\includegraphics[width=8.6cm]{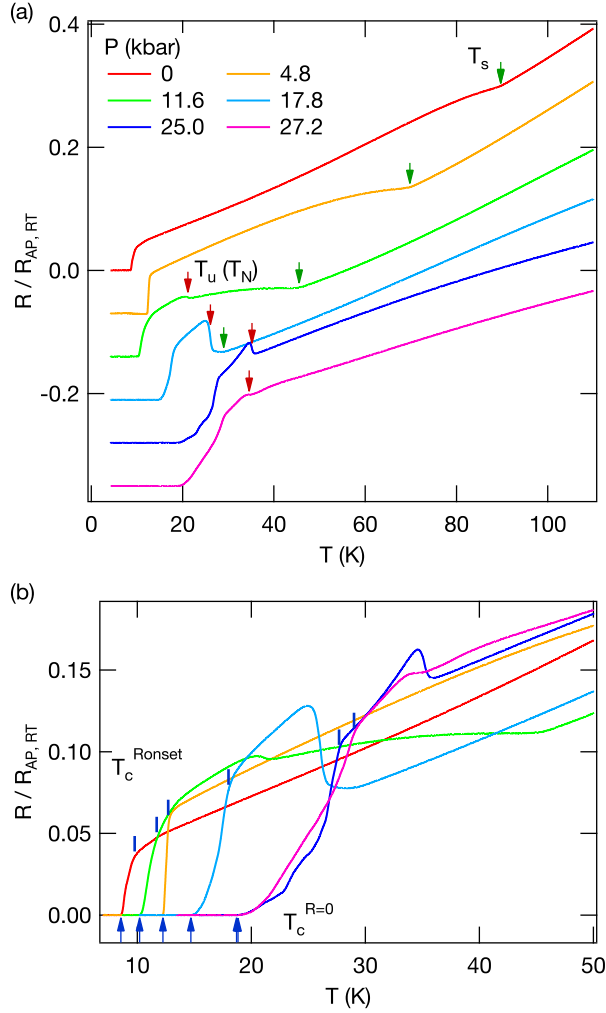}
\caption{\label{RvsT}(Color online).  (a) In-plane resistance of FeSe at selected pressures normalized by the ambient-pressure room-temperature value as a function of temperature (curves are vertically shifted for clarity) and (b) a blowup of a low-temperature part (curves are not shifted).  The structural transition temperature $T_s$ [green (online) down-pointing arrows in (a)] is defined by a positive peak of d$^2R$/d$T^2$.  The unknown but most likely antiferromagnetic transition temperature $T_u$ [red (online) down-pointing arrows in (a)] is defined by a negative peak of d$R$/d$T$.  The superconducting transition temperatures $T_c^{R = 0}$ [up-pointing arrows in (b)] and $T_c^{Ronset}$  (short vertical bars) are defined by the zero resistance and a negative peak of d$^2R$/d$T^2$, respectively.}   
\end{figure}

\begin{figure}
\includegraphics[width=8.4cm]{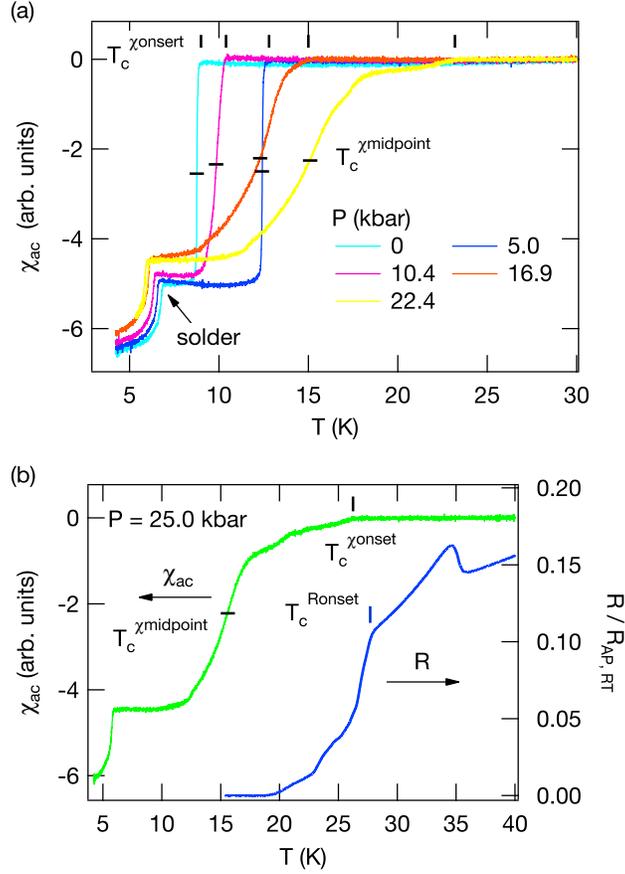}
\caption{\label{Xac}(Color online).  (a) AC magnetic susceptibility of FeSe at selected pressures as a function of temperature.  The ac excitation field is approximately parallel to the $ab$ plane.  The vertical bars indicate the superconducting transition temperatures $T_c^{\chi onset}$ determined from the onset of the diamagnetism, while the horizontal bars the transition midpoint $T_c^{\chi midpoint}$ (we have arbitrarily assumed that the signals at $T$ = 7 K correspond to the full shielding).  The curves have not been corrected for the background variation of the empty pickup coil, and the anomalies at $T$ = 6 -- 7 K are due to the superconducting transition of solder used for wiring.    (b) Comparison between the temperature dependences of ac magnetic susceptibility and resistance at $P$ = 25.0 kbar.  $T_c^{\chi onset}$ and $T_c^{Ronset}$ are indicated by vertical bars.  $T_c^{\chi midpoint}$ is also indicated.}   
\end{figure}

\begin{figure}
\includegraphics[width=8.6cm]{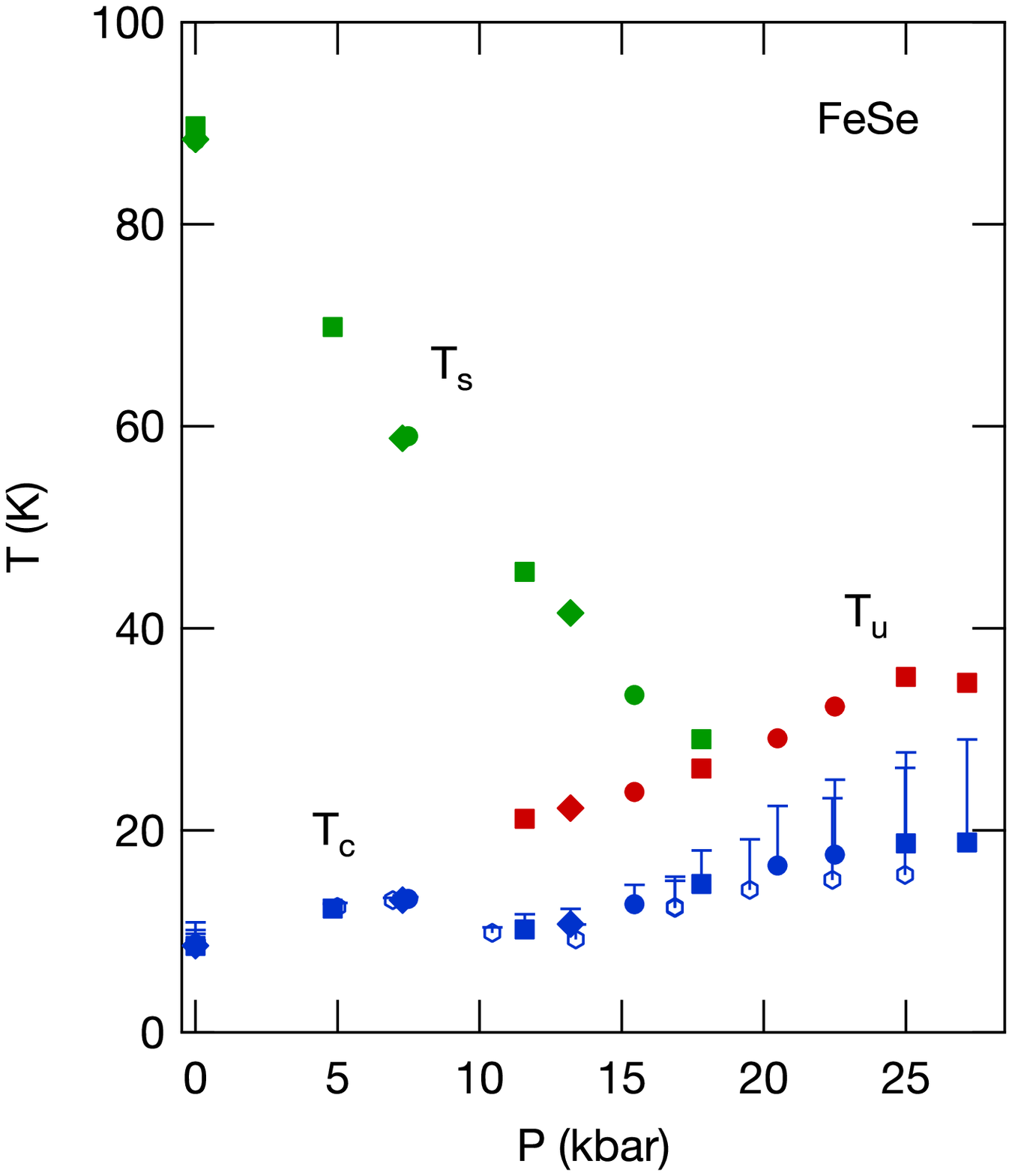}
\caption{\label{phase}(Color online).  High-pressure phase diagram of FeSe. Structural transition at $T_s$ [green (online) symbols], unknown but most likely antiferromagnetic transition at $T_u$  [red (online) symbols], and superconducting transition at $T_c$ [blue (online) symbols].  For $T_c$, the solid and hollow symbols indicate the zero-resistance temperatures $T_c^{R = 0}$ and the midpoint temperatures of the ac magnetic susceptibility $T_c^{\chi midpoint}$, respectively.  The vertical bars show the transition widths, and the horizontal bars indicate the onset temperatures $T_c^{Ronset}$ and $T_c^{\chi onset}$ .  Different symbol shapes correspond to different samples.}   
\end{figure}

Figure~\ref{RvsT} shows selected temperature dependences of resistance.
At ambient pressure ($P$ = 0 kbar), the anomaly associated with the structural phase transition is seen at $T_s$ = 89.7 K [green down-pointing arrow in (a)].
The transition temperature $T_s$ is rapidly suppressed as pressure is applied, reaching $T_s$ = 29.0 K at $P$ = 17.8 kbar.
The anomaly is no longer seen above this pressure.
Another anomaly appears at $T_u$ = 21.2 K at $P$ = 11.6 kbar [red down-pointing arrow in (a), also see (b)].
The anomaly is much clearer at $P$ = 17.8 kbar.
The transition temperature $T_u$ increases with pressure up to 25.0 kbar, but appears to decrease slightly at $P$ = 27.2 kbar, the highest pressure of the experiment.
Although the resistance increase at $T_u$ is fairly sudden, no clear hysteresis was found: the hysteresis width, if any, would be very small, less than $\sim$0.1 K. 
The superconducting transition temperature [zero-resistance criterion, up-pointing arrows in (b)] initially increases from $T_c^{R = 0}$ = 8.6 K at $P$ = 0 kbar to 12.2 K at $P$ = 4.8 kbar (d$T_c$/d$P$ = 0.76 K/kbar), and the transition becomes sharper.
The transition temperature, however, decreases to 10.2 K at $P$ = 11.6 kbar, and the transition becomes broader.
As pressure is increased further, the transition temperature increases again, but the transition width broadens further.
For the evaluation of the transition width, we have defined the onset temperature $T_c^{Ronset}$ as a temperature where d$^2R$/d$T^2$ shows a negative peak [short vertical bars in (b)]. 
Between the two highest pressures, 25.0 and 27.2 kbar, the zero-resistance temperatures $T_c^{R = 0}$ are nearly the same, though the onset temperature still shows an increase.
Resistance measurements performed on two more samples gave similar observations.

Figure~\ref{Xac}(a) shows selected temperature dependences of ac magnetic susceptibility.
Note that the anomalies at about 6 -- 7 K are due to the superconducting transition of solder used for wiring.
As pressure is applied, the superconducting transition temperature $T_c^{\chi onset}$ (onset criterion, indicated by vertical bars) initially increases from 9 K at $P$ = 0 kbar to 12.8 K at 5.0 kbar, but decreases to 10.4 K at 10.4 kbar, where the transition is broader.
As pressure is further increased, $T_c^{\chi onset}$ increases again, but the transition broadens further.
At $P$ = 22.4 kbar, the diamagnetism appears at $T_c^{\chi onset}$ = 23.2 K but remains weak until $\sim$18 K, below which it grows faster.
Figure~\ref{Xac}(b) compares the temperature dependences of ac magnetic susceptibility and resistance at $P$ = 25.0 kbar.
The diamagnetism appears at $T_c^{\chi onset}$ = 26.2 K, which is close to the onset temperature of the resistance drop $T_c^{Ronset}$ = 27.7 K, though most of the growth of the diamagnetism occurs below the zero-resistance temperature $T_c^{R=0}$ = 18.7 K. 
Considering the broad transitions at high pressures, we have also defined the transition midpoints $T_c^{\chi midpoint}$ (horizontal bars in Fig.~\ref{Xac}), which might serve as a better index of the bulk transition temperature.
We note that the size of the diamagnetic signal remains nearly the same as pressure is applied, indicating bulk superconductivity up to the highest pressure of 25.0 kbar.
We did not detect anomalies corresponding to $T_s$ or $T_u$, most likely because of the low sensitivity of the present setup.

Figure~\ref{phase} shows the high-pressure phase diagram of FeSe determined from the present measurements.
The superconducting transition temperature initially increases, but decreases above $\sim$8 kbar and then increases again, resulting in a local minimum at $\sim$12 kbar.
This is fully consistent with previous reports.\cite{Bendele10PRL, Bendele12PRB, Miyoshi14JPSJ}
Furthermore, the observation that the superconducting volume fraction hardly changes in the investigated pressure range is also consistent with those reports.
The structural transition temperature $T_s$ is quickly suppressed by the application of pressure.
This is consistent with a previous report,\cite{Miyoshi14JPSJ} though the suppression rate observed in the present study is faster.
The fate of the structural transition above $P$ = 17.8 kbar is not clear and will be discussed later.

The resistance anomaly at $T_u$ probably corresponds to the antiferromagnetic order observed in the high-pressure $\mu$SR study.\cite{Bendele10PRL, Bendele12PRB}
According to Bendele \textit{et al.},\cite{Bendele10PRL, Bendele12PRB} a finite magnetic volume fraction appears at $P$ = 8 kbar with $T_N$ = 17 K, but it does not reach 100\% as $T \to 0$ at this pressure.
As pressure is increased above 8 kbar, the magnetic volume fraction and $T_N$ increase while $T_c$ decreases. 
At 12 kbar, the volume fraction reaches 100\% (as $T \to 0$) and $T_c$ reaches a local minimum.
Above 12 kbar, both $T_N$ and $T_c$ increase with pressure up to 24 kbar, the highest pressure of Ref.~\onlinecite{Bendele12PRB}.
The reported behavior of $T_N$ is basically consistent with the behavior of $T_u$ observed in the present study.
The increase in the resistance at $T_u$ indicates that the Fermi surface is partially gapped by the antiferromagnetic order.
The quantitative agreement between $T_N$ and $T_u$ is however not very good: $T_N$ = 55 K at $P$ = 24 kbar (Ref.~\onlinecite{Bendele12PRB}) vs. $T_u$ = 35.2 K at 25.0 kbar.
We note that $T_N$ in Refs.~\onlinecite{Bendele10PRL, Bendele12PRB} is defined as the temperature where the magnetic volume fraction becomes nonzero.
However, the volume fraction increases rather slowly with decreasing temperature: e.g. the magnetic volume fraction at $P$ = 24 kbar reaches 100\% only below 30 K.
This indicates that $T_N$ in the sample of Ref.~\onlinecite{Bendele12PRB} is distributed between 55 and $\sim$ 30 K, which may explain the discrepancy between the reported $T_N$ and the present $T_u$.
The present $T_u$, especially above 15 kbar, is close to the temperature where the volume fraction reaches 70\% in Ref.~\onlinecite{Bendele12PRB}: the 70\% line starts at $\sim$12 kbar at $T$ = 0 and reaches $\sim$35 K at 24 kbar.

Interestingly, the pressure dependence of $T_u$ and that of $T_c$ are roughly parallel: both transition temperatures increase with pressure from $\sim$12 to 25 kbar and then approximately level off.
This is very different from behavior expected in usual quantum-critical-point (QCP) scenarios, which assume that the superconducting dome is centered at the QCP, i.e., $T_c$ is maximum at the QCP.
%In some cases, $T_c$ might be dipped in the vicinity of the QCP, as predicted by some theories of ferromagnetic superconductors for example \cite{Fay80PRB, Roussev01PRB}.
%This could explain the minimum of $T_c$ at $P \sim 12$ kbar, where $T_u$ starts to appear, but still the rough parallel between $T_c(P)$ and $T_u(P)$ on the ordered side of the phase diagram is not explained.
It would be helpful to extend the pressure range to see how $T_c$ and $T_u$ evolve above 27 kbar.
It would also be interesting to see how theories claiming competition between different spin fluctuations in FeSe could explain our observation.\cite{Lischner15PRB, Glasbrenner15condmat} 

The width of the superconducting transition considerably broadens as $T_u$ increases with pressure.
At high pressures, the onset of diamagnetism approximately coincides with the onset of resistance drop [Figs.~\ref{Xac}(b) and \ref{phase}].
This is quite unusual.
Usually the diamagnetism develops after the zero resistance is achieved, since the zero resistance requires only a one-dimensional current path, which can be as thin as possible, but for the diamagnetism to be observed some volume must be shielded.
In the present case, isolated superconducting regions appear below $T_c^{\chi onset}$ but their connections are prevented by some reason until the temperature becomes much lower than $T_c^{R=0}$.
This likely reflects the competition between the superconductivity and magnetism despite the simultaneous increase of $T_c$ and $T_u$.
It is interesting to note that a similar observation has been reported for FeSe thin films: the onset of the diamagnetism approximately coincides with the onset of the resistance drop for three and four unit-cell films, for which $T_c \sim45$ K.\cite{Deng14PRB}
ARPES papers suggest that FeSe thin films may order antiferromagnetically with $T_N > 100$ K.\cite{Tan13NatMat, He13NatMat}

Finally, we ask what is the fate of the structural transition $T_s$ above $P$ = 17.8 kbar.
High-pressure structural studies suggest that the low temperature structure remains orthorhombic up to $\sim$75 kbar.\cite{Margadonna09PRB, Kumar10JPCB, Uhoya12EPL}
One scenario compatible with this is as follows:
The $T_s(P)$ transition line merges with the $T_u(P)$ line at some pressure, above which a stripe-type antiferromagnetic order and orthorhombic distortion occur simultaneously at $T_u$, as is the case with BaFe$_2$As$_2$, for example.
On the other hand, those structural data were obtained for mixed-phase samples and might not reflect the intrinsic phase diagram.\cite{Margadonna09PRB, Kumar10JPCB, Uhoya12EPL} 
Thus the following scenario, among others, is also worthy of consideration:
The structural transition temperature $T_s$ is suppressed down to zero (continuously or in a first-order fashion) at some pressure, above which the structure remains tetragonal as $T \to 0$.
We note that the $T_u(P)$ line shows no clear kink marking the point where the $T_s$ and $T_u$ transition lines meet.
This might suggest that the two orders are rather decoupled, as has been suggested by ambient-pressure studies,\cite{Bohmer13PRB, Baek14nmat, Bohmer15PRL} and hence might support this scenario.
In this case, the antiferromagnetic order would not be a stripe-type one but would have to be compatible with the tetragonal structure as has been suggested in Ref.~\onlinecite{Bohmer15PRL}.
Although Ref.~\onlinecite{Bendele12PRB} proposes stripe-type antiferromagnetic ordering under high pressure based on the $\mu$SR data, neutron diffraction measurements have failed to confirm it. 
Clearly, high-pressure structural measurements on high-quality single crystals are desired.

In summary, we have found an anomaly most likely corresponding to the pressure-induced antiferromagnetic transition in our resistance measurements on FeSe.
No clear hysteresis associated with this anomaly was observed.
The obtained phase diagram shows intriguing relations between the three orders:
The antiferromagnetic and superconducting transition temperatures show analogous pressure dependences as if the two orders are cooperative.
Nevertheless, the superconducting transition width broadens in the pressure region where the antiferromagnetism coexists as if the two orders compete.
The structural transition line almost meets the antiferromagnetic one at 17.8 kbar, the highest pressure where the transition was observed.
The antiferromagnetic transition line smoothly evolves in the neighboring pressure range, which might suggest that the two orders are fairly independent of each other.
The fate of the structural transition above this pressure deserves further investigations.

\begin{acknowledgments}
This work has been supported by Japan-Germany Research Cooperative Program, KAKENHI from JSPS, Project No. 56393598 from DAAD, the ¡ÈTopological Quantum Phenomena¡É (No. 25103713) KAKENHI on Innovative Areas from MEXT of Japan, and JSPS KAKENHI Grant Number 26400373.
\end{acknowledgments}

\end{document}